\documentclass[pre,reprint,superscriptaddress]{revtex4-1}
\usepackage{graphicx}
\usepackage{xcolor}
\usepackage{amsmath}
\usepackage{hyperref}

\begin{document}

\title{Spatial patterning of force centers controls folding pathways of active elastic networks}

\author{Debjyoti Majumdar}
\email{debjyoti@post.bgu.ac.il \\ Current address: RIKEN Center for Biosystems Dynamics Research, Kobe, Japan}
\affiliation{Department of Environmental Physics, Jacob Blaustein Institutes for Desert Research,\\
Ben-Gurion University of the Negev, Sede Boqer Campus 84990, Israel}

\date{\today}

\begin{abstract}
We study the effect of the spatial distribution of active force dipoles on the folding pathways and mechanical stability of rigid-elastic networks using Langevin dynamics simulations. While it has been shown in Majumdar et al., J. Chem. Phys. 163, 114902 (2025) that a sharp collapse transition is evident in triangular (elastic) bead-spring networks under the action of contractile (or extensile) force dipoles distributed randomly across the network, here, we show that when the spatial distribution is correlated, e.g., like a patch in the center (``active core'' model) or a band-like distribution along the periphery (``active periphery'' model), the network undergoes only a partial decrease in size even at large forces, thereby showing an enhanced mechanical stability just from a spatial rearrangement of the active dipoles. Further, an active periphery network shows higher mechanical stability initially, for a range of forces, beyond which the active core network becomes more stable. Deformation in the network becomes irreversible beyond a threshold force, which depends on the type of distribution; for a uniform distribution of active dipoles, the irreversibility threshold almost coincides with the critical collapse point, it decreases for the active core system, and is decreased further for the active periphery system. It is shown that irreversibility arises due to plastic deformations in the form of crease formation which is not reversible even after the force is turned off or reversed. The folding pathways depend weakly on the temporal stochasticity of the active links, but are highly sensitive to any defects (missing bonds) in the network. Our findings, therefore, suggest active force localization (or delocalization) as a prime method to dynamically alter the mechanical stability and reversibility of the underlying elastic network.
\end{abstract}

\maketitle

\section{Introduction}

Folding is a fundamental process that is crucial for the proper functioning of many natural and artificially engineered systems. For instance, in biology, protein folding dictates whether a polypeptide chain reaches its functional three-dimensional structure, while misfolding can lead to severe pathologies \cite{dobson2003}. Chromatin, on the other hand, requires a hierarchical folding before it is packed tightly into the cell nucleus \cite{fraser2015}. Sequential irreversible folding is required during morphogenesis for organ formation in 3D \cite{teranishi2024}.  At larger scales, folding mechanisms are central to the operation of deployable self-automated robotic sheets \cite{zhou2025} and programmable origami-inspired metamaterials \cite{stern2017}. Despite the diversity of contexts in which folding is found, there is a similarity in the underlying mechanism in that it involves the interplay of internal stresses and geometry. While generally only the final folded state matters most, understanding the pathway followed and the stress propagation can also be important to understand possible kinetic trapping into metastable states. In this paper, we aim to study the folding and mechanical stability of elastic networks representing elastic sheets under the action of active force dipoles.

In recent years, there has been a surge of interest in the dynamics of elastic networks driven by active forces \cite{gov2007, broedersz2011, das2012, sheinman2012, alvarado2013, alvarado2017, gnesotto2019, shivers2019, arzash2021, arzash2023, dasbiswas2023}. These forces are generated internally, e.g., by molecular motors \cite{howard2001}, contractile units, or embedded actuators \cite{baconnier2022}, rather than applied externally. Such active stresses are known to fundamentally alter the material's response, producing emergent phenomena such as rotation,  contractility \cite{singh2024, majumdar2025}, or large-scale pattern formation \cite{bois2011, kumarV2014, rombouts2023}. It can even enhance the rigidity of the network, even if the system is poised well below the rigidity percolation point \cite{majumdar2025}.  For most cases, bead-spring networks were used to represent elastic media; for example, triangular networks of various dilutions have been employed to study biopolymer networks, such as the actomyosin network \cite{sheinman2012,das2012,alvarado2013,broedersz2014,gnesotto2019,dasbiswas2023,singh2024}. Similar network models have also been used in designing mechanical networks with controlled response to external stimulus, both globally  \cite{goodrich2015},  and locally \cite{rocks2017,yan2017}. From an engineering perspective, these model mechanical networks have been instrumental in studying fracture propagation \cite{sanner2025}.

Recently, it has been shown that rigid triangular bead-spring networks can undergo a sharp collapse transition under the action of both contractile or extensile active force dipoles, when the dipole links are distributed uniformly across the network \cite{majumdar2025}. Further, it was demonstrated that while active dipole links result in a collapse transition, conversely, it is the same dipoles that rigidify the network by effectively increasing the number of constraints. The effect of any spatial correlation in the distribution of active links was not considered in that study. In biological systems,  however,  processes often require the cooperative action of these active units, requiring them to self-organize in clusters through a mechanochemical feedback. For instance, in actomyosin networks, myosin-II often clusters into mini-filaments, which act as contractile units \cite{howard2001}. These clusters localize preferentially at regions of high actin filament density, crosslinks, or nodes of mechanical tension \cite{prost2015}. In reconstituted in vitro systems, myosin-II can spontaneously self-organize into clusters, even in isotropic networks  \cite{kohler2011}. It has been shown that the same actin filament crosslinkers either enhance or inhibit the contractility of a network, depending on the organization of actin within the network \cite{ennomani2016}.  {Structural correlation has been shown to lower the rigidity percolation point to a lower volume fraction \cite{zhang2019}.} However, a systematic study of the effect of spatial segregation of these active dipoles is still unknown. 

 \begin{figure}[t]
\centering
\hspace{-3.5cm}(a)~~~~~~~~~~~~~~~~~~~~~~~~~~~~~~~~(b)\\
\includegraphics[width=.45\linewidth]{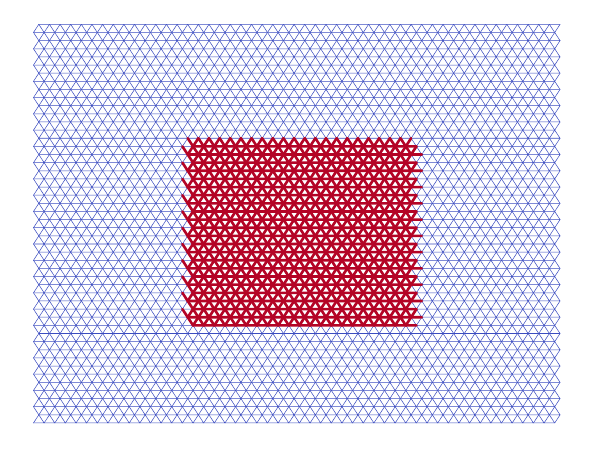}
\includegraphics[width=.45\linewidth]{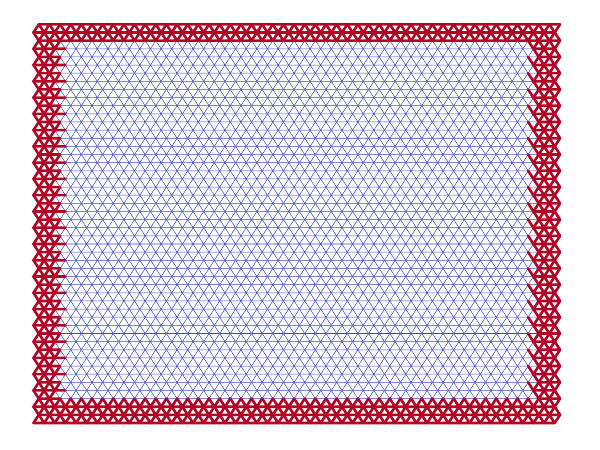}\\
\hspace{-3.5cm}(c)~~~~~~~~~~~~~~~~~~~~~~~~~~~~~~~~(d)\\
\includegraphics[width=.45\linewidth]{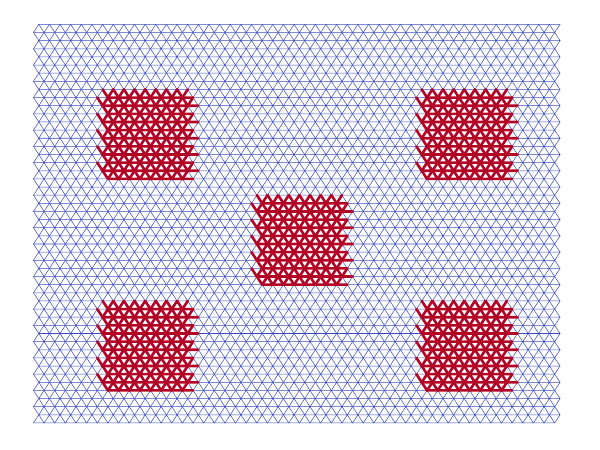}
\includegraphics[width=.45\linewidth]{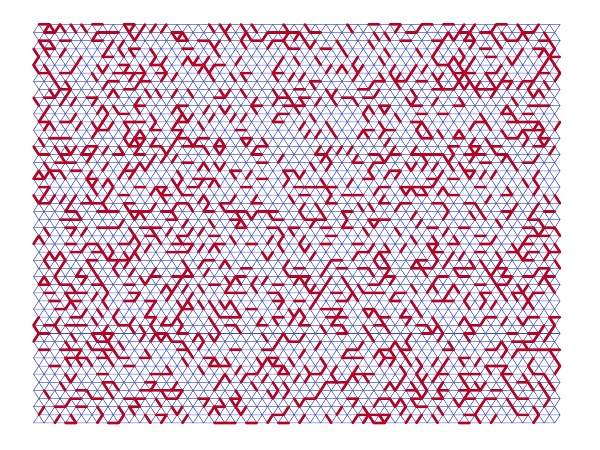}

\caption{Triangular lattices with different active dipole link distributions: (a) a patch like distribution of active dipoles at the center of the network; (b) active dipoles distributed like a band along the periphery; (c) active dipoles distributed like small patches across the network; (d) uniformly distributed active dipoles. All four types of distribution contain about $\phi=0.2$ fraction of active dipole links highlighted in red.}
\label{fig_schematics}
\end{figure}

To investigate how the spatial distribution of active force dipoles leads to differential mechanical properties, we consider two different models of active dipoles' spatial distribution on the triangular lattice, and compare them to that of the uniform case for which an incipient sharp collapse transition is already known to exist \cite{majumdar2025}. We start with the case where the active units are clustered at the center of the elastic medium. Thereafter, we consider the problem where the same fraction of links is distributed like a band along the periphery of the network. A physical manifestation of this model could be the actomyosin network in the cells, which rigidifies the cell boundary. Both models show that upon a clustering of the active units, the network is less prone to losing its mechanical stability under the action of these active forces, unlike the uniform distribution, where the system collapses to a length scale comparable to the bond length (in the absence of hydrodynamic and volume exclusion interaction) through a sharp dynamical collapse transition. Further, we could identify two distinct force regimes in which the two considered distributions of active links can make the network alternatively more mechanically stable than the other. 

The rest of the paper is organized in the following manner: in Sec \ref{model}(A), we define the elastic network model, in Sec \ref{model}(B), model for the active dipolar forces, in Sec \ref{model}(C), model for the spatial patterning of these dipole links and simulation methods in Sec \ref{model}(D). In Sec \ref{observables}, we discuss the observables of interest relevant to our study. In Sec. \ref{results}(A), we discuss the findings for the active core model, the active periphery model in Sec. \ref{results}(B),  hysteresis and reversibility under deformation in  Sec. \ref{results}(C), and the dependence of the folding pathways on the temporal stochasticity and its sensitivity to any defects present in the network in Sec. \ref{results}(D). Finally, we conclude with a discussion on the applicability of our results to different systems and possible future outlooks in Sec. \ref{conclusion}.

\section{Model and methods \label{model}}
\subsection{Elastic network}
We model the elastic network as a triangular lattice where the lattice sites represent beads which are connected to the neighboring beads via harmonic springs. A triangular lattice bead-spring network naturally represents a rigid system, since the mean coordination number $(z)$ is higher than the critical $z_c$ required by the Maxwell criteria for rigidity using constraint counting \cite{comm1, maxwell1864}. Therefore, any infinitesimal deformation will be met with restoring forces bringing the system back to its initial state as the force is removed. The lattice is of linear dimension $L=50$ consisting of $N=2500$ nodes and $7301$ bonds. The beads are connected via harmonic springs of rest length $b=1$ and spring constant $K=m\omega_0^2$, where $m$ is the bead mass and $\omega_0$ is the spring self-frequency. The  dynamics for each bead is performed following the Hamiltonian
\begin{equation}
\mathcal{H} = \frac{1}{2}K \sum_{\langle ij \rangle} (\vec r_i - \vec r_j - b\hat r_{ij})^2 
\end{equation}
where $\langle ij \rangle$ represent neighboring connected nodes. Since our system lacks translational invariance, owing to the presence of localized active links, and also because dynamical shape transitions form a part of our study we do not apply any periodicity across the boundaries. As a result, the average mean coordination number is $\langle z \rangle=5.84$ slightly less than that of an infinite lattice for which $\langle z \rangle_{\infty}=6$ . 

\subsection{Active forces}

\begin{figure}[t]
\centering
\includegraphics[width=.5\linewidth]{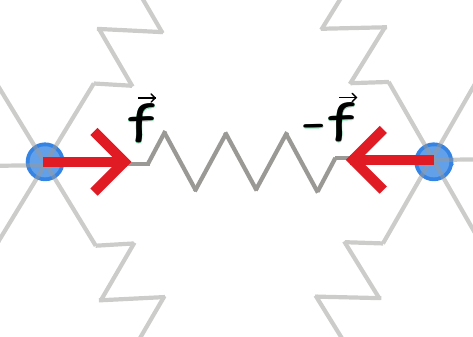}
\caption{Schematic diagram depicting a contractile force dipole acting on two neighboring nodes connected with a spring.}
\label{fig_active_force_schematic}
\end{figure}

We model the active forces as ``force dipoles'' which act along the bond connecting two nodes. Therefore, the force field is given by
\begin{equation}
\vec F(\vec r, t)=\sum_i \vec f_j(t) \left[ \delta(\vec r - \vec r_j) - \delta(\vec r - \vec r_j - \vec \epsilon_j) \right],
\end{equation}
where $\vec \epsilon_j (\mid \vec \epsilon_j \mid = b)$ is a vector directed along one of its neighbors. The force dipoles can be contractile ($f>0$) in which case the forces act inwards, bringing the participating nodes closer together [Fig. \ref{fig_active_force_schematic}], or extensile $(f<0)$ in which case the nodes are pushed away from each other. In either case, since equal and opposite forces act simultaneously, there is no net momentum transfer, and, therefore, no drift in the network center of mass. Note that our force dipoles are not point-like; rather, they are of the size of the instantaneous bond length that they occupy.

Initially, at time $t=0$, the force dipoles are independently in an ON (or OFF) state with a probability $P$ (or $1-P$). The force dipoles switch between ON $(f)$ and OFF $(f=0)$ states independently of each other in the manner of a ``telegraphic'' process. The active forces are temporally correlated with a decorrelation time $\tau$. The intermittent switching of states is to be associated, e.g., with the ``attachment'' and ``detachment'' of myosin motors within the actomyosin network. The number of times an active dipole switches between different states  within a time interval is Poisson-distributed. For details on switching statistics, refer to Appendix \ref{appendix_active_force}. 

For simplicity, we have not included any long-range hydrodynamic interaction, excluded volume interaction, or bond bending energy. We believe that the results will remain equally valid even with those interactions included, although the stability and critical points might change.

\subsection{Patterned activity in the network}
To model the patterned distribution of active components on the network we mainly consider two distinct situations of the active link distribution; first, a square patch of active links is placed at the center of the elastic network which we call as the ``active core'' model. Second, the active links are distributed along the periphery like a band, which we refer to as the ``active periphery'' model. The fundamental difference between these models lies in the way in which the passive bulk  will be subjected to active forcing; an active core pulls the matrix from inside, while an active periphery squeezes the matrix from outside. Additionally, we also consider an extension of the active core model, where equal size patches are located across the network. However, we found its effect  to be equivalent to that of a single active core.  We will also consider the uniform distribution of active links recently discussed in  Ref. \cite{majumdar2025}. For brevity we will refer to the uniform distribution of active links as the ``uniform distribution'', henceforth. 

\subsection{Simulation Methods}

We perform Langevin dynamics simulations of the bead-spring network solving the following equation of motion 
\begin{equation}
m\ddot{\vec r}_i =  -\gamma \dot{\vec r}_i +  \vec F_i^e + \vec F_i^a 
\end{equation}
 in the over-damped limit $(\ddot{\vec r}\rightarrow 0)$, where $m$ is the bead mass,  $\gamma$ is the friction coefficient,  $\vec r_i$ is the position of the $i$th node,  and $\vec F^e_i$ and $\vec F^a_i$ are the conservative spring forces and  non-conservative active forces acting on the node $i$, respectively. We set $\gamma=1$ and $m=1$ throughout our simulations. To integrate we use the two-point Euler method. We use a unit system where position $\vec r$ is mentioned in units of bond length $b$, active force $\vec f$ in units of $m\omega_0^2b$, and time $t$ and $\tau$ are in units of $\tau_0=\gamma/(m\omega_0^2)$. An integration time step of $\delta t=10^{-2}$ has been used. See Appendix \ref{appendix_int_scheme} for more discussions.

\section{Observables \label{observables}}

The mechanical stability of the network against the active forces is attributed to the steady-state time-averaged $R_g$. That $R_g$ can be used as a measure for the mechanical stability of the system follows from our previous work Majumdar et al. \cite{majumdar2025}, where we have shown that for marginally stable networks in 2D, such as the square lattice, the $R_g$ always decreases under the action of dipole forces even for tiny forces, albeit slowly. On the other hand, full or sparsely diluted triangular networks that are well above the 2D isostatic point reach a steady state even for sufficiently large but sub-critical forces, with a well-defined mean $R_g$, and thereby give a collapse transition above a threshold force. The sharpness, however, decreases as we dilute the system towards the isostatic point \cite{majumdar2025}. Therefore, we can safely assume that a higher $R_g$ at the same active force corresponds to a network of higher mechanical stability in the sense that it can sustain stress and maintain its structure at the same time. The radius of gyration is calculated on the fly using 
\begin{equation}
R_g = \sqrt{\frac{1}{N} \sum_{j} (\vec{r}_j-\vec{r}_{\text{cm}})^2}
\label{eq_rg_main}
\end{equation}
where $\vec{r}_{\text{cm}}=\frac{1}{N}\sum_j \vec r_j$ is the center of mass of the full system. Further, to identify the individual contribution of the active and passive parts to the network size, we split the $R_g$ up into active $R_{g}^a$ and passive $R_{g}^p$ components given by 
\begin{eqnarray}
R_g^2 &=& \frac{1}{N} \left[ \sum_{j=1}^n (\vec{r}_j^{~a}-\vec{r}_{\text{cm}})^2 +  \sum_{j=n+1}^N (\vec{r}_j^{~p}-\vec{r}_{\text{cm}})^2 \right]  \\
&=& \frac{N_a}{N}(R_g^a)^2 + \frac{N_p}{N}(R_g^p)^2
\label{eq_rg_components}
\end{eqnarray}
made up of the active $R_g^a = \sqrt{\frac{1}{N_a} \sum_{i\in a} ( \vec r_i  - \vec r_{\text{cm}})^2}$, and  passive $R_g^p=\sqrt{\frac{1}{N_p} \sum_{i\in p} ( \vec r_i  - \vec r_{\text{cm}})^2}$ parts calculated using the active $(N_a)$ and passive $(N_p)$ nodes, respectively. Notice that, for both $R_{g}^a$ and $R_{g}^p$, we choose the same center of mass of the whole system, since the system is essentially symmetric about it. 

Apart from $R_g$, we also look at the stress $(\sigma)$ on individual network nodes, where the stress on node `$i$' is calculated using the formula

\begin{equation}
\sigma^i_{\alpha\beta} = -\sum_{\langle ij\rangle} \vec f^{~ij}_{\alpha} . \vec r^{~ij}_{\beta}
\label{eq_sigma_mean}
\end{equation}

where the sum $\langle ij\rangle$ is taken over the neighboring connected nodes, $\alpha,\beta$ are the $x,y$ components, $\vec f_{ij}$ is the force acting on the node $i$ due to node $j$, and $\vec r_{ij}$ is the displacement vector from the node $j$ to node $i$. Of our interest are the diagonal normal stress components of the stress tensor $\sigma_{xx}$ or $\sigma_{yy}$, which tell us how much the network is being compressed - for a negative value, or stretched apart - for a positive value, along each direction. To see stress accumulation, we calculate the mean of the two components, i.e., $\sigma_{mean}=\frac{1}{2}(\sigma_{xx}+\sigma_{yy})$.

\begin{figure}[t]
\centering
\hspace{-3.5cm}(a) ~~~~~~~~~~~~~~~~~~~~~~~~~~~~~~~~~~(b)\\
\includegraphics[width=.49\linewidth]{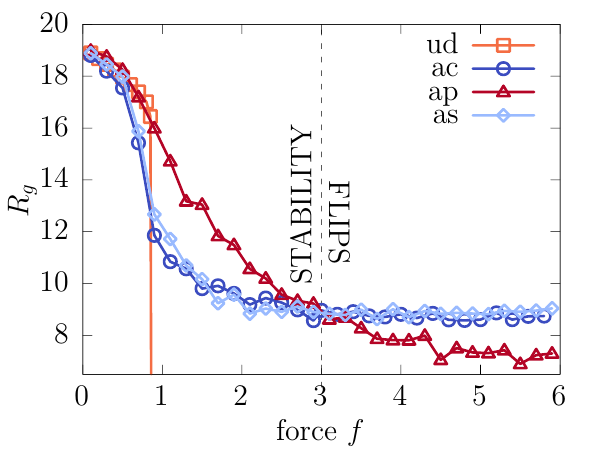}
\includegraphics[width=.49\linewidth]{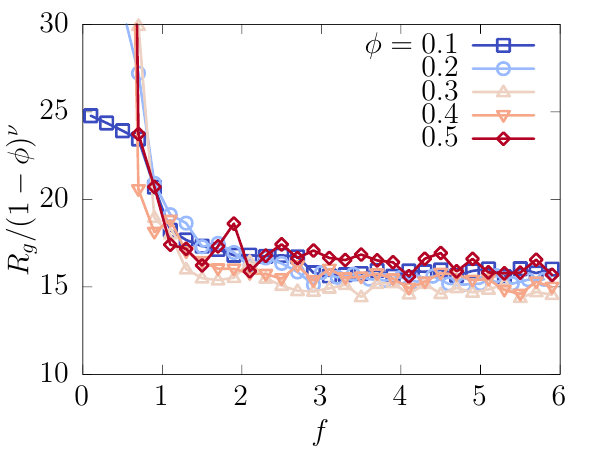}\\\hspace{-3.5cm}(c)~~~~~~~~~~~~~~~~~~~~~~~~~~~~~~~~~~(d)\\
\includegraphics[width=.49\linewidth]{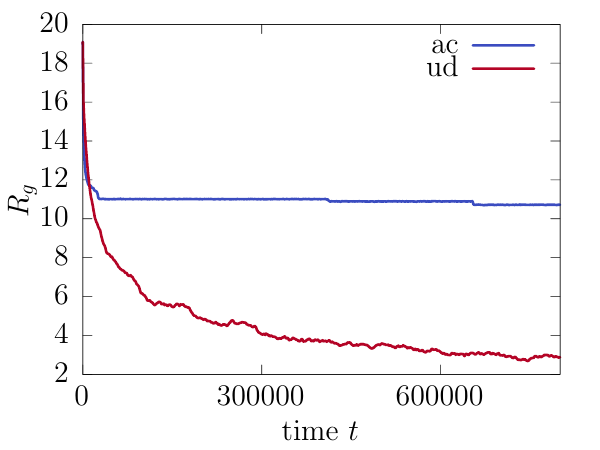}
\includegraphics[width=.49\linewidth]{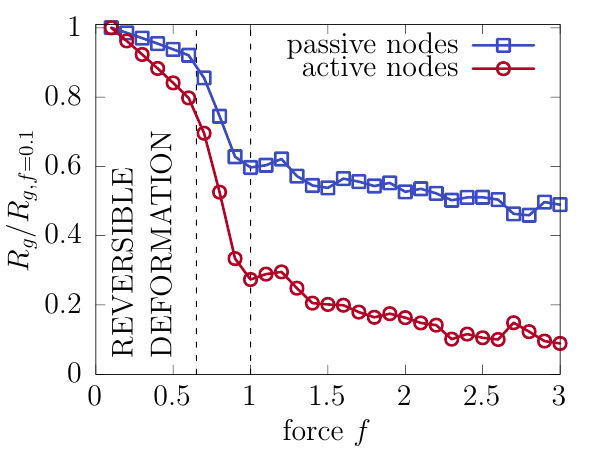}
\caption{ (a) Time-averaged steady state radius of gyration $R_g$ plotted as a function of force for the active core (ac), active periphery (ap) and active patches (as) networks with $\phi = 0.2$ fraction of dipole links. For comparison we have also shown the data for the uniform distribution (ud), previously reported in Ref. \cite{majumdar2025}. (b) For forces $f>1$, $R_g$ scales as $(1-\phi)^{\nu}$ with $\nu=2.5$ for different patch sizes containing fraction of dipole links $\phi=0.1-0.5$. (c) Comparison between the relaxation time to reach the steady state for ud and ac at $f=1$. (d) Relative decrease of the active and passive $R_g$ components of the ac network. The radius of gyration $R_g$ has been normalised by the $R_g$ at $f=0.1$. The dashed lines correspond to $f=0.65$ and  $f=1$.}
\label{fig_rg_active_core}
\end{figure}

\section{Results \label{results}}

\subsection{Active core passive surrounding}

To begin with, we consider a patch like density of contractile active force dipoles in the center of the network representing an elastic system with an active core [Fig. \ref{fig_schematics}(a)]. The dipole links are chosen within a bounded area such that the fraction of dipole links is close to $\phi=0.2$ (exactly $\phi=0.202$) to compare with the $\phi=0.2$ result in Ref. \cite{majumdar2025}. Force decorrelation time used is $\tau=1$, which is kept unchanged for other simulations as well. Initially ($t=0$) the force dipoles are in an ON or OFF state with a probability of $P=1/2$. Since there is no temperature in the system $(T=0)$, the dynamics come solely from the stochastic switching between ON and OFF states of the force dipoles and the resultant restoring force from the network springs.

As the system evolves, a steady state is reached where the overall system size reduces and the time-averaged $R_g$ do not changes with time anymore except small fluctuations about the mean value. To calculate $R_g$, we start time averaging only after the system is deep inside the steady-state. With the increase of the active force amplitude, we do observe a decrease in the $R_g$ value, characteristic of the contractile force dipoles [Fig. \ref{fig_rg_active_core}]. To see the dynamics refer to multimedia file Mov1.mp4 available online. However, unlike the sharp collapse transition at $f_c\approx 0.9$ observed in Ref. \cite{majumdar2025} for $\phi=0.2$ fraction of dipole links distributed uniformly across the network, we do not find a similar transition to a complete collapse state corresponding to $R_g\approx b=1$ at $f\geq 1$. Instead, $R_g$ remains finite even after substantial decrease, for high value of forces $f>1$ [Fig. \ref{fig_rg_active_core}(a)]. For $f>1$, the steady state $R_g$ decreases with increasing patch size or equivalently the fraction of dipole links $(\phi)$ and follows a scaling $R_g \sim (1-\phi)^{\nu}$ with $\nu=2.5$ [Fig. \ref{fig_rg_active_core}(b)]. The exponent $\nu$ comprises a $1/d$ factor from the  passive bulk excluding the collapsed active core, and a non-trivial contribution due to the deformation under the specific pattern of dipole distribution [see Appendix \ref{appendix_scaling_exponents}]. As compared to the uniform distribution, the active core relaxes about  $10^3$ orders of magnitude faster at $f=1$ [Fig. \ref{fig_rg_active_core}(c)]. This could be because of critical slow down in the uniform system since $f=1$ is in the vicinity of the collapse critical point, while there is no such criticality in the active core system. For non-critical forces, however,  reaching steady state is generally fast for the uniform case also. For the active core, we do see rare step like small drops in $R_g$ at long intervals of time which majorly comes from the passive part of the $R_g$. However, such drops arising from inward displacement of the passive nodes, due to the continuous activity of the core, are small enough $(<b)$, so as not to change the steady state $R_g$ significantly. For further increase of force up to $f=6$ the system seems to be stable in the steady state at around $R_g=9$ without any further decrease in $R_g$. This absence of a complete collapse even at large forces can be apprehended as an increased mechanical stability in the network, arising simply from a spatial rearrangement of the active dipoles.

\begin{figure}[t]
\centering
\hspace{-3.5cm}(a) ~~~~~~~~~~~~~~~~~~~~~~~~~~~~~~~~~~(b)\\
\includegraphics[width=.49\linewidth]{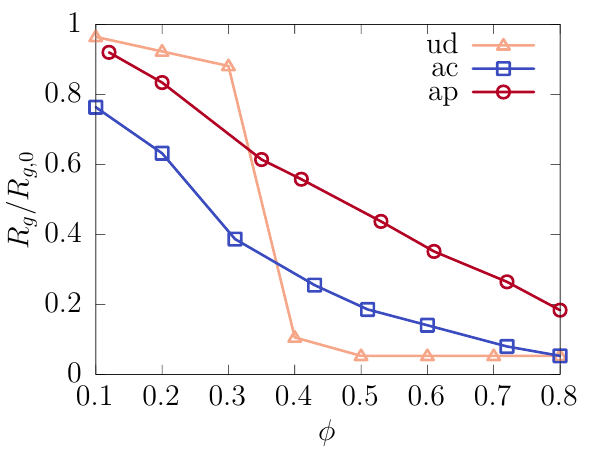}
\includegraphics[width=.49\linewidth]{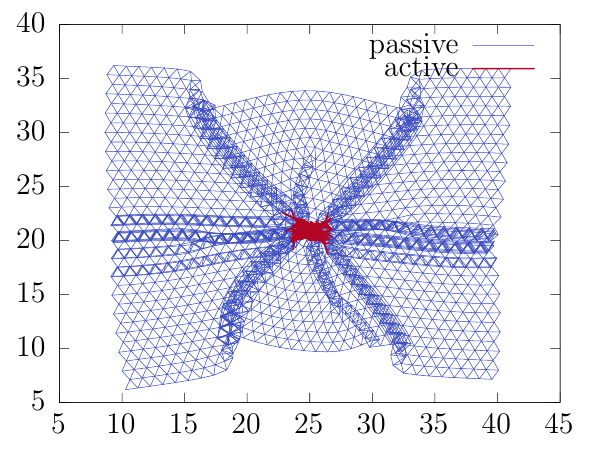}\\
\hspace{-3.5cm}(c) ~~~~~~~~~~~~~~~~~~~~~~~~~~~~~~~~~~(d)\\
\includegraphics[width=.49\linewidth]{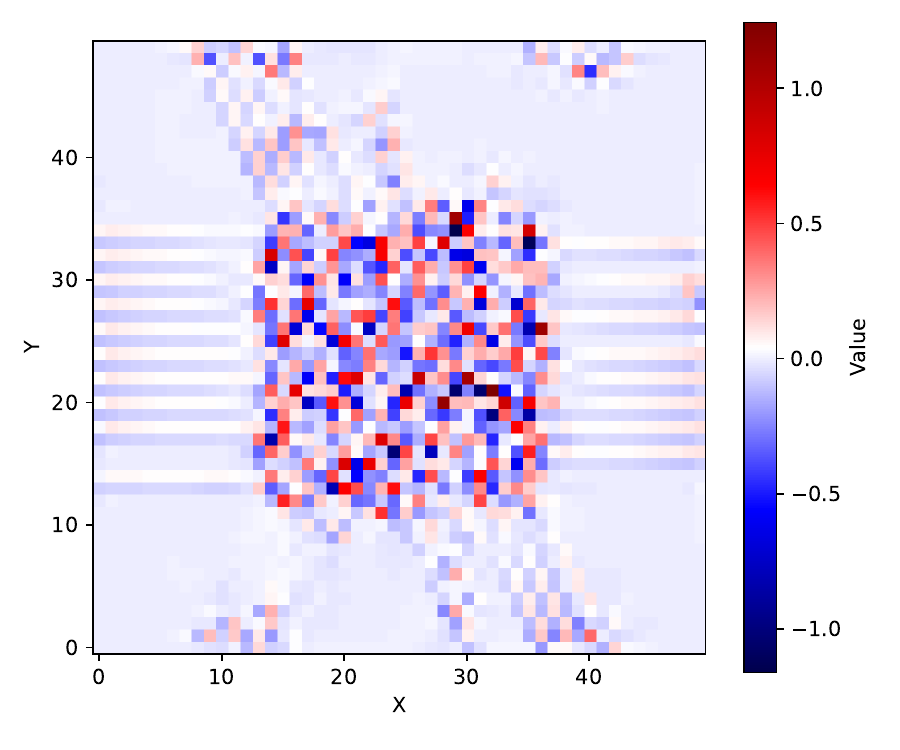}
\includegraphics[width=.49\linewidth]{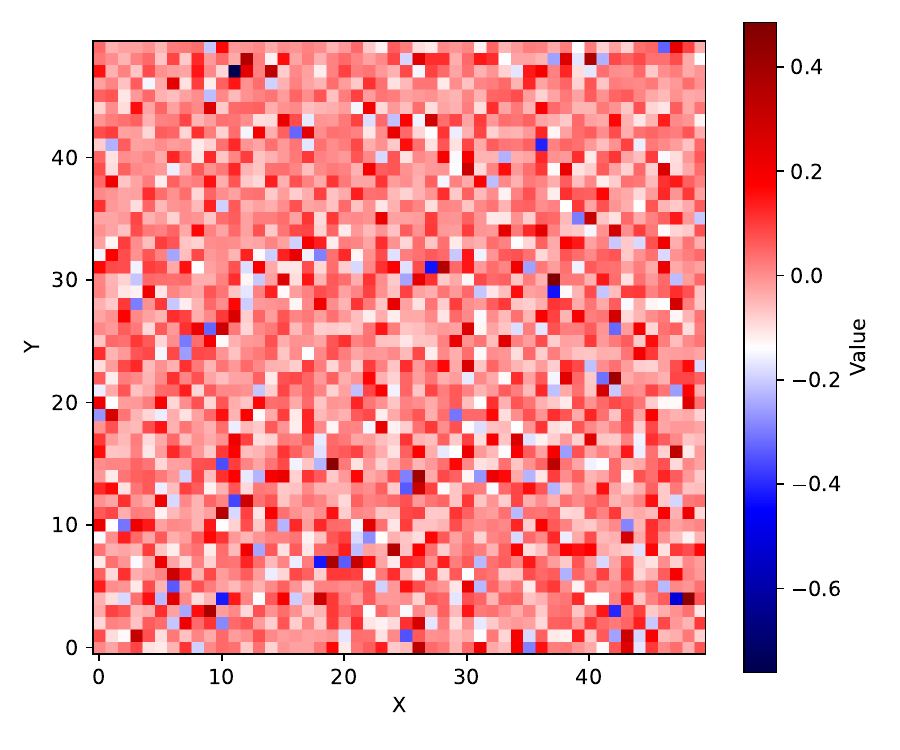}
\caption{ (a) $R_g$ as a function of $\phi$ at a constant fixed active force amplitude $f=1$ for active core (ac) and active periphery (ap). For uniform distribution (ud) the force is fixed at $f=0.65$. (b) Near steady state snapshot for the active core system at a force amplitude $f=3$. The passive and active parts of the network are colored in blue and red, respectively. (c-d) Heatmap plot of the steady-state mean stress $\sigma_{mean}$ including both active and passive components of forces and calculated using Eq. \ref{eq_sigma_mean} for the ac network in (c) and ud network in (d).}
\label{fig_stress_active_core}
\end{figure}

To find out how the individual active and passive network components shrink, we calculate $R_g$ of the active and passive nodes separately using Eq. \ref{eq_rg_components}. In Fig. \ref{fig_rg_active_core}(d), we have plotted $R_g/R_{g,f=0.1}$ to visualise the different regimes of shrinking. We found three distinct regimes where $R_g$ scales differently with the force: an initial regime for $f\lesssim 0.6$, where only the active nodes show a significant decrease; second, for $0.6<f\lesssim 1$ with the fastest decrease in $R_g$ for both active and passive nodes, and finally for $f>1$ where the decrease in $R_g$ slows down with $R_g^a$ reducing slowly towards zero as $f$ increases, while $R_g^p$ remains finite.  To see  how $R_g \rightarrow 0$ as $\phi \rightarrow 1$,  we gradually increase the fraction of dipole links and plot $R_g$ as a function of $\phi$ at a constant force $f=1$  in Fig. \ref{fig_stress_active_core}(a). We found that a collapsed phase is reached only gradually without any sharp change in $R_g$ when $\phi$ is varied. On the contrary, for an uniform distribution a sharp collapse by varying $\phi$ is possible around $\phi=0.35$ for $f=0.65$. This shows that $\phi$, just like the force $f$, can act as a critical variable only for an appropriate distribution of active dipoles.

The absence of a complete collapse results from how the spatial distribution of active force centers change the way stresses are transmitted across the network. When uniformly distributed, contractile forces from force dipoles spread from the source to other parts of the network by branching into force trees at each consecutive nodes with the subsequent force dipoles along the path assisting in force transmission. On the other hand, for a clustered patch of active dipoles, of size sufficiently smaller than the system, and surrounded by a passive elastic bulk, the forces do not propagate uniformly throughout the network, consequently a ``butterfly-like'' structure, buckled along the edges, forms due to non-symmetric distribution of stress. In Fig. \ref{fig_stress_active_core}(b), we have shown the steady-state structure at $f=3$. Further, the passive bulk surrounding the active core, redistributes and balance stresses, making the system stable even at high force amplitudes. 

To look at the stress distribution across the network, we calculate $\sigma_{mean}$ using Eq. \ref{eq_sigma_mean} on the individual nodes and plot it in a heatmap type plot in Fig. \ref{fig_stress_active_core}(c) for active core and Fig. \ref{fig_stress_active_core}(d) for uniform distribution, where each pixel corresponds to one node. For active core we choose $f=3$, while for the uniform distribution we choose $f=0.8<f_c$. Clearly, for the active core system we can see that stress accumulation happens only on the parts where the network form folds or creases, besides the active core itself. The nodes in the active region carry the majority of stresses due to contribution from the active components.  For uniform distribution, however, stress is uniformly distributed across the network including parts corresponding to $\sigma_{mean}\approx 0$ [Fig. \ref{fig_stress_active_core}(d)]. Noticeably, for the active core, there are regions in the diagonal direction where stress propagation is minimal even though they are near to the active core dipoles, relative to regions which are far away but still carry significant amount of stress, e.g, parallel to the x-direction around the center of the network. This shows that active dipole patches can be used to tactically channel stress along a specific path on the network crucial for achieving a desired folding or even to transmit mechanical signal. 

To explore what happens if the clustered patch dynamically breaks into many smaller patches, we break the single patch into 5 smaller patches distributed across the network [Fig. \ref{fig_schematics}(c)]  while keeping the combined fraction of dipole links $\phi\approx 0.2$. Interestingly, $R_g$ for different forces overlaps with that of the active core datapoints [Fig. \ref{fig_rg_active_core}(a)]. This suggests that the contributions of the individual patches in contracting the network may combine in an approximately additive manner as long as the individual patch sizes $>b$. To verify this, we further introduce anisotropy in the patch sizes, but $R_g$ shows little change. However, further analysis is needed to confirm this interpretation. Overall, breaking the active region into smaller chunks have little effect on the  mechanical stability, since, the $R_g$ remains essentially same, however, the folding pattern has changed. This highlights a potential route for dynamical rearrangements that alter morphology while preserving key mechanical properties such as rigidity.

\subsection{Active periphery passive core}

\begin{figure}[t]
\centering
\hspace{-3.5cm}(a) ~~~~~~~~~~~~~~~~~~~~~~~~~~~~~~~~~~(b)\\
\includegraphics[width=.49\linewidth]{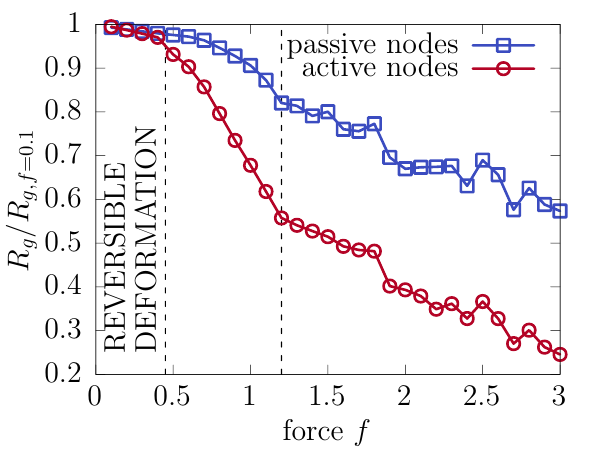}
\includegraphics[width=.49\linewidth]{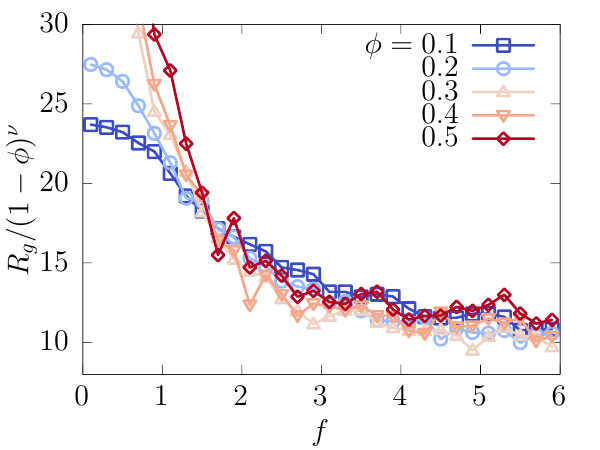}\\
\hspace{-3.5cm}(c) ~~~~~~~~~~~~~~~~~~~~~~~~~~~~~~~~~~(d)\\
\includegraphics[width=.49\linewidth]{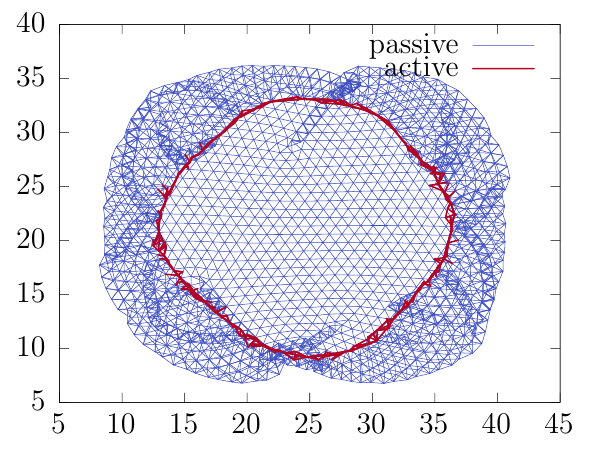}
\includegraphics[width=.49\linewidth]{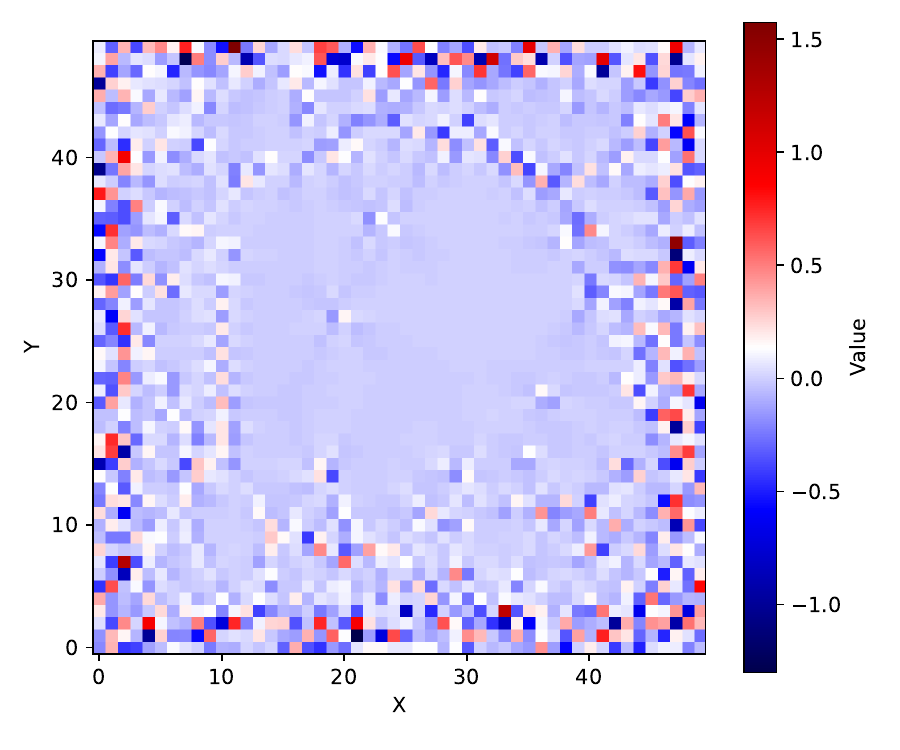}
\caption{(a) Relative decrease of the active and passive components of the active periphery (ap) network. The radius of gyration $R_g$ has been normalised by the $R_g$ at $f=0.1$. The dashed lines correspond to $f=0.45$ and  $f=1.2$. (b) Active periphery network size scaling in the high force regime $(f>1)$ with fraction of dipole links $\phi$. $R_g$ scales as $(1-\phi)^{\nu}$ with $\nu=1.75$. (c) Steady state snapshot for the active core system at a force amplitude $f=3$. The passive and active parts of the network are colored in blue and red, respectively. (d) Heatmap plot of the steady-state mean stress $\sigma_{mean}$ at force $f=3$, including both active and passive components of the forces, calculated using Eq. \ref{eq_sigma_mean}.}
\label{fig_rg_active_peri_scaling}
\end{figure}

Next, we consider the opposite of the active core problem, where the outer periphery is active and the inner bulk is passive. The fraction of dipole links is kept same around $\phi=0.2$ (exact value $\phi=0.195$). In this case also, the $R_g$ shows only a partial decrease even for high forces [Fig. \ref{fig_rg_active_core}(a)]. Interestingly, we find that for forces $f<3$, the active periphery with higher $R_g$  is more stable compared to the active core, while for forces $f>3$ the active core  system possesses higher mechanical stability [Fig. \ref{fig_rg_active_core}(a)]. The crossing between the two regime happens somewhere around $f\approx 3$ where both the systems are equally stable with similar $R_g$, although the individual configurations are different. For forces $f>2$, the network size in the steady state scales with $\phi$ (or equivalently the width of the active periphery) as $R_g\sim (1-\phi)^\nu$ with $\nu=1.75$ [Fig. \ref{fig_rg_active_peri_scaling}](b). Looking at the steady state configurations and the corresponding movies, we found the pathways followed during the shrinking of the network is different than that for the active core system (see animation Mov2.mp4 available online).  For instance, here, the periphery starts folding on itself crossing the bulk towards the center of the system while minimizing the peripheral size [Fig. \ref{fig_rg_active_peri_scaling}(c)]. There is negligible stress propagation towards the passive core due to the absence of any excluded volume among the nodes, so that the inner nodes do not resist to the outer peripheral contraction [Fig. \ref{fig_rg_active_peri_scaling}(c)]. On the contrary, active core dipoles induces long-range peripheral wrinkling and buckling, albeit in specific directions.

Similar to the active core case, here also, a $\phi$ induced collapse transition seems to be absent, rather a gradual crossover takes place to a collapsed state as we increase $\phi$ keeping the force amplitude fixed at $f=1$ [Fig. \ref{fig_stress_active_core}(a)]. When comparing the relative decrease in size of the active and passive components, we found three scaling regimes of $R_g$  [see Fig. \ref{fig_rg_active_peri_scaling}(a)]. However, the different components shrink slower than the corresponding components in the active core system, suggesting that pulling-in is perhaps a more effective way to change (reduce) the system size than pushing from outside, whereby the force transmission happens more effectively across the network and for the same amount of resources used (here, fraction of dipole links).

\subsection{Reversibility under deformation \label{section_reversibility}}

It is often necessary to restore a system to its initial state, e.g, by gradual removal of the applied field, which is the active force here. The restoring dynamics may simply retrace the path followed when the field was applied, or it can follow a different route showing a phase coexistence like in hysteresis of spin models \cite{newman1999}.  Therefore, we study the $R_g$ vs. $\vec f$ curve under gradual increase, removal and reversal of active forces. To begin with, we set a maximum force amplitude of $f_{\text{max}}=1.2$, noting that $f_c=0.9$ is the collapse critical force for the uniform distribution. Then, starting from $f=0$ we slowly increase the force amplitude upto $f_{\text{max}}$, then reduce it to $-f_{\text{max}}$, with $f<0$ representing extensile dipole forces, and then again increase it back to $f_{\text{max}}$. Each variation of the force amplitude is done at a step of $\delta f=0.1$. At every new value of the force, the system runs for $2\times 10^6$ steps, and the averaging over $R_g$ starts from $1.8\times 10^6$ steps onwards. We found that during decreasing force, the system does not trace the same path it initially took when the force was increased, {thereby showing dependence on the history of states visited before}. Moreover, even after completing one cycle of loading and unloading, which includes the force to go from contractile to extensile, the $R_g$ still decreases, denoting that the system always evolves towards an absorbing phase, which is the completely collapsed phase.

\begin{figure}[t]
\centering
\hspace{-3.5cm}(a) ~~~~~~~~~~~~~~~~~~~~~~~~~~~~~~~~~~(b)\\
\includegraphics[width=.49\linewidth]{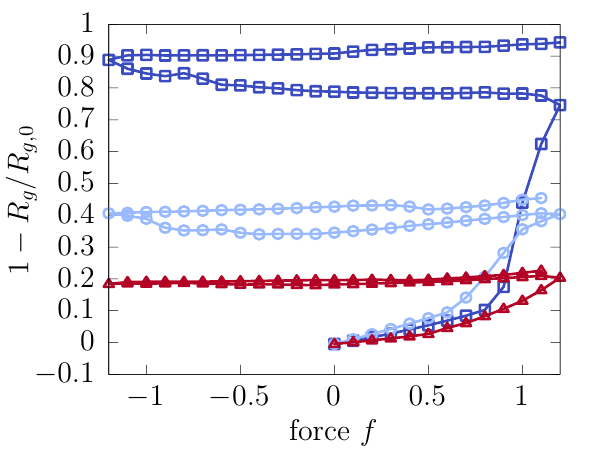}
\includegraphics[width=.49\linewidth]{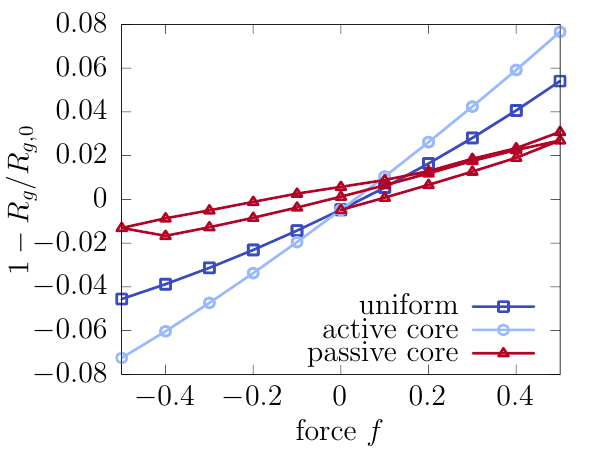}\\
\hspace{-3.5cm}(c) ~~~~~~~~~~~~~~~~~~~~~~~~~~~~~~~~~~(d)\\
\includegraphics[width=.49\linewidth]{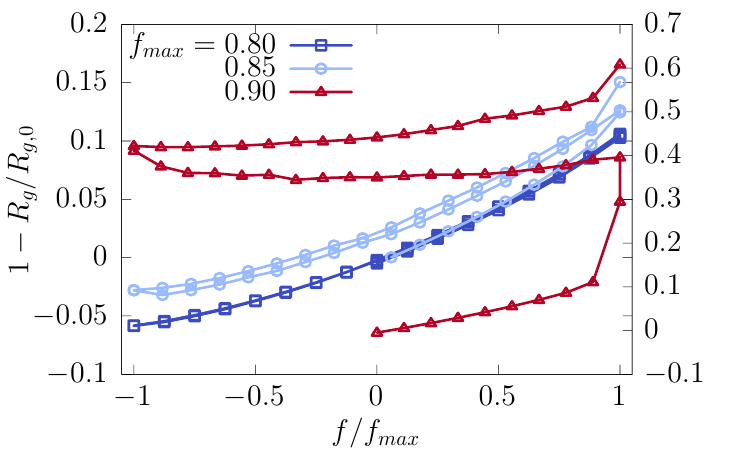}
\includegraphics[width=.49\linewidth]{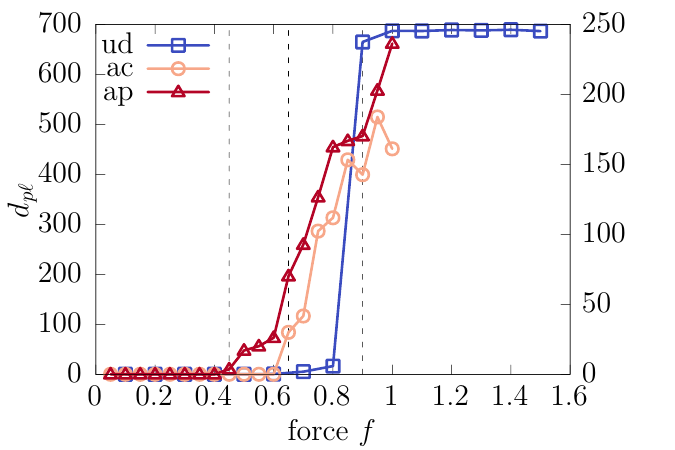}
\caption{(a)-(b) In the y-axis we plot $1-R_g/R_{g,0}$, where $R_{g,0}$ is the $R_g$ at $f=0$, and force along x-axis. Positive (negative) force values correspond to contractile (extensile) dipoles. The maximum force amplitude is set to $f_{\text{max}}=1.2$ in (a), and $f_{\text{max}}=0.5$ in (b). The protocol followed is $f=0\rightarrow f_{\text{max}}\rightarrow -f_{\text{max}} \rightarrow f_{\text{max}}$. In panel (b), the unbifurcated straight lines for uniform distribution and active core are  reversible deformations, while irreversibility appears as a  bifurcated line for the active periphery. Panel (a) uses the same color coding as (b). (c) Same as (a) but only for uniformly distributed active dipoles at $f_{\text{max}}=0.8, 0.85$ and $0.9$. The left (right) vertical y-axis corresponds to $f_{\text{max}}=0.8$ and $0.85~(0.9)$. The horizontal x-axis has been normalized by $f_{\text{max}}$. (d) Plastic deformation measured using $d_{p\ell}$ from Eq. \ref{eq_dpl}. Left (right) vertical axis corresponds to uniform distribution (active core and active periphery). The  vertical dashed lines correspond to $f=0.45$, $0.65$ and $0.9$.}
\label{fig_reversibility}
\end{figure}

This motivates us further to explore if the system traces the same path during loading and unloading for smaller forces, along with a threshold point for the onset of irreversibility. Interestingly, we found that the force thresholds for the system deformation to become irreversible are different for the three cases considered. For the uniform distribution, the irreversibility point lies very near to the collapse transition point $f_c=0.9$. To verify this, we studied for $f_{\text{max}}=0.8,~0.85$ and $0.9$ [Fig. \ref{fig_reversibility}(c)]. Since they are close to the critical point, we further increase the time of equilibration at each new force value by an order of magnitude up to $10^7$, and the averaging starts from $9\times 10^6$. We found that while for $f_{\text{max}}=0.8$ the system is perfectly reversible, small irreversibility starts appearing for consecutive cycles from $f_{\text{max}}=0.85$, which becomes large at the critical point $f_{\text{max}}=0.9$. For the active core case, however, the point lies somewhere within $f=0.6$ - $0.7$, and for the active periphery, it lies in between $f=0.4$ - $0.5$.  When we compare these regimes with that of the change in the active and passive $R_g$ components Fig. \ref{fig_rg_active_core}(d), we found that the scaling of $R_g$ actually starts showing a substantial change from the irreversibility point onwards. 

The irreversibility under removal or reversal of force, signifies a buildup of permanent (plastic) microscopic changes in the network. Since topological changes are not allowed in our model, such plastic deformations can only arise due to the formation of creases, whereby, connected neighboring nodes change their spatial ordering, i.e., right neighbor becomes left neighbor and vice-versa, without changing the network topology. To quantify such plastic deformations, we add-up the magnitude of displacements given by
\begin{align}
d_{p\ell} =
\frac{1}{2}\sum_{i=1}^N  \sum_{ j>i} \Big[ \mid \Delta  x_{ij} \mid \left(1-\text{sgn}(\Delta  x_{ij} \Delta  x_{ij}^0) \right) \delta_{j,i+1} \nonumber \\ 
+ \mid \Delta  y_{ij} \mid \left(1-\text{sgn}(\Delta  y_{ij} \Delta  y_{ij}^0) \right) (1-\delta_{j,i+1}) \Big] 
\label{eq_dpl}
\end{align}
where the first sum is over all the $N$ nodes, the second sum $\langle i,j \rangle$ is over the neighbors, $\Delta x_{ij}^0 = x_j^0 - x_i^0$ ($\Delta x_{ij}$) is the displacement in the undeformed (deformed) network, $\text{sgn}(x)$ gives the sign of $x$, $\delta_{i,j}$ is the Kronecker delta, and plot it as a function of the applied force in Fig. \ref{fig_reversibility}(d). Clearly, the sharp rise in $d_{p\ell}$ coincides with $f_c=0.9$ for uniform distribution, $f=0.65$ for active core network, and $f=0.45$ for active periphery. All three $f$ values marking a significant rise in $d_{p\ell}$, are in good agreement with that of the irreversibility thresholds estimated from Fig. \ref{fig_reversibility}(a)-(c), thereby proving that irreversibility is a consequence of plastic deformations.

Because, for a uniform distribution of active dipoles, active stresses are distributed more evenly across the network, the system resists extreme structural reconfigurations, until the global collapse point is reached. Hence, irreversibility and collapse are nearly coincident resulting in the highest degree of reversibility. Active core, on the other hand, leads to stress localization, thereby irreversible plastic deformation occurs more readily. The core collapses even before the whole network reaches any collapse point. The system thus enters an irreversible regime much earlier due to local plastic changes.  Active periphery behaves even more fragilely; active edges undergo irreversible boundary reconfigurations at very low forces due to a lack of structural support from the outside, leading to edge instabilities at very small forces. Similar irreversible transitions are also evident, e.g., during epithelial folding, which is induced by F-actin accumulating into a bracket-like structure via a mechanosensitive response of cells \cite{teranishi2024}.

\subsection{Sensitivity to temporal stochasticity and defects in network}

Next, we ask, how sensitive are the folding pathways and therefore the final state reached to the temporal stochastic switching of the active forces in different samples but with the same spatial distribution ?  
To characterize  we perform a procrustes type analysis where we first find the mean shape from 30 samples, and then find the deviation of each sample from this mean shape using the formula
\begin{equation}
\mathcal{O} =\sqrt{ \frac{1}{N}\sum_{i=1}^N \mid \vec r_i^\alpha - \vec r_i^m \mid^2}
\end{equation}
where $i=1$ to $N$ runs over the nodes of the network, $\alpha$ over different sample networks and $\vec r_i^m$ represents the mean configuration given by $\vec r^m_i = \frac{1}{N_s}\sum_{\alpha} \vec r_i$, where $N_s$ is the number of samples used. However, no aligning operations such as rotation or reflection was made since the force dipole patch do not cause any drift in the center of mass or rotations.  Note that the lowest possible value for the variable is $\mathcal{O}=0$ for two exactly same configurations, while the upper value is unbounded.

\begin{figure}[t]
\centering
\hspace{-3.5cm}(a) ~~~~~~~~~~~~~~~~~~~~~~~~~~~~~~~(b)\\
\includegraphics[width=.4\linewidth]{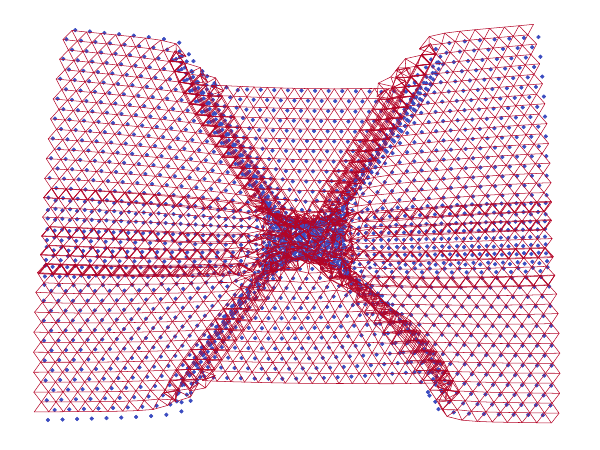}
\includegraphics[width=.58\linewidth]{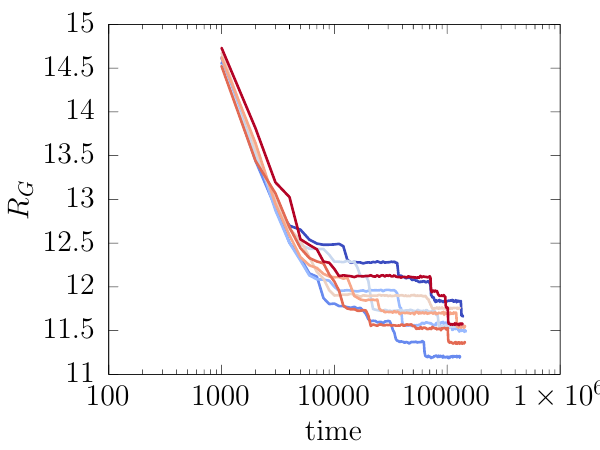 }\\
\hspace{-6cm}(c)\\
\includegraphics[width=\linewidth]{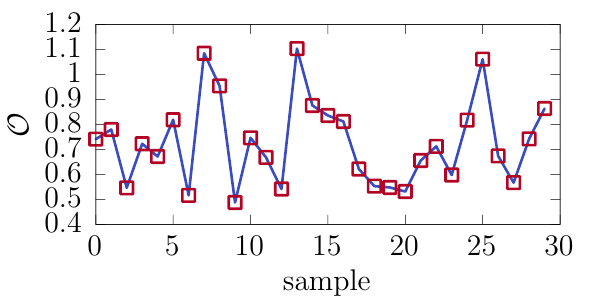}
\caption{(a) Nodes of a folded network (in blue dots) averaged over 30 independent configurations, and an actual folded network (in red) superimposed on it at $f=1$. (b) $R_g$ time evolution for 8 different samples with same spatial active link distribution but different temporal sequence of switching between ON and OFF states. (c) Similarity between each sample with the average folded state.}
\label{fig_temporal_similarity}
\end{figure}

We analyse only the active core system; for low forces, such as $f=0.3$, it is found that with the same spatial distribution of active links but with different switching timings, we have the same final state reached by all samples. 
Even at high forces, such as $f=1$, the steady state pattern formed has high degree of similarity among all the samples. For visualization, we give an overlapping plot of the mean configuration and one of the samples in Fig. \ref{fig_temporal_similarity}(a). In Fig. \ref{fig_temporal_similarity}(b), we show the time evolution of $R_g$ for $8$ different samples. The deviation of each sample from the mean sample is small $\mathcal{O}\approx b$ as shown in Fig. \ref{fig_temporal_similarity}(c). This opens up a new  possibility of controlled folding of elastic networks (or sheets) specifically patterned to obtain a desired fold without needing to worry about the the temporal sequence of stochasticity. Additionally, it means that the macroscopic folding dynamics is mainly dominated by the spatial position of the active links, and depends only weakly on the exact timings of active sources changing their states. In other words, the temporal randomness averages out and {on the time scales the sheet relaxes, the stochastic switching behaves like an averaged (effective) active field (e.g. $\langle f(\vec r, t)\rangle \approx P f)$, where the dynamics is governed by that mean field force}. In terms of energy landscape, this also hints at the existence of a strong attractor or  basin of attraction such that the system and its energy landscape funnel many different forcing histories into the same trajectory and a final minimum.

\begin{figure}[t]
\centering
\hspace{-8cm}(a) \\
\includegraphics[width=\linewidth]{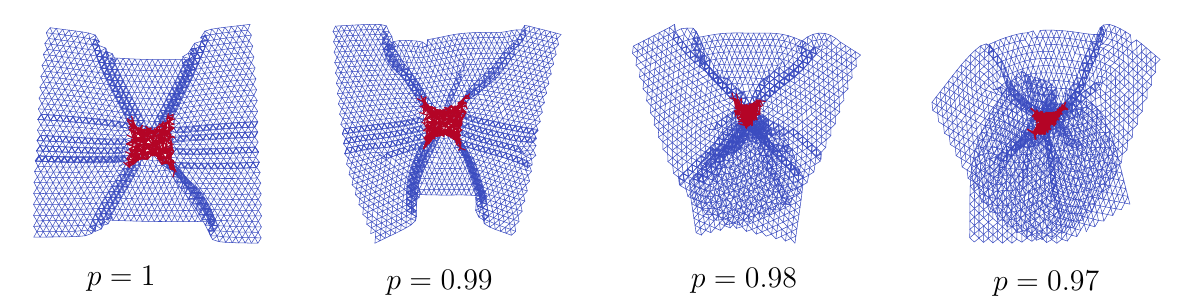}\\
\hspace{-8cm}(b) \\
\includegraphics[width=\linewidth]{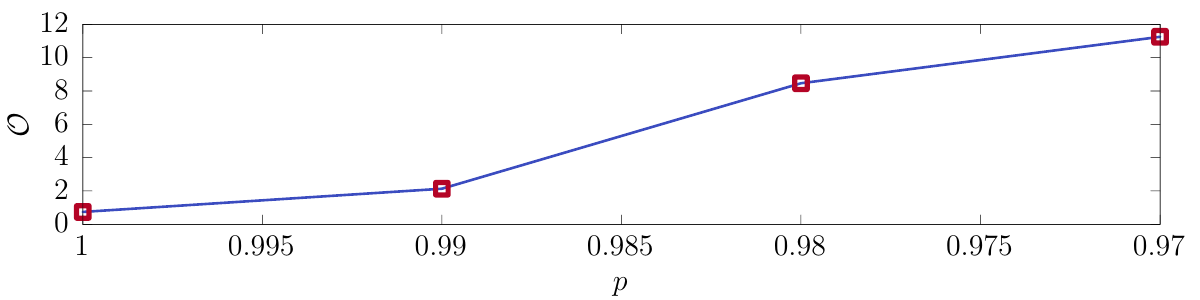}
\caption{(a) Final folded state for different dilution values $p=1,0.99,0.98$ and $0.97$ where $p$ is the probability with which a bond exists, at force $f=1$ and for same fraction of active links $\phi\approx 0.2$ up to second decimal place. (b) Deviation $(\mathcal{O})$ of the configurations from the mean folded structure for $p=1$.}
\label{fig_defect}
\end{figure}

For targeted folding, it is also important to understand how robust it is to the presence of any defect in its functional components. For our system, defect can be realised, e.g., as the absence of few bonds in the network. This motivates us to ask how sensitive is the folding of the network to the final target state if a fraction of bonds are removed from the network.  Here, also we will use the active core as our test case since both wrinkling and folding of segments are distinctly visible. To introduce defect we randomly dilute some bonds across the network which can also include active links. For the uniform case it is already known that diluting the triangular network makes the collapse transition smoother towards a second order transition  \cite{majumdar2025}.  In a similar line, we found the folding to also depend strongly on even a minuscule amount of defect resulting into a catastrophic change in the final folded structure as we dilute even as small as $2$ percent of  the bonds, i.e., $p=0.98$ where $p$ is the probability with which a bond is present. While the random dilution does not significantly change the fraction of active links (e.g., for $p=0.97$ it is $\phi=0.204$), the final structure is drastically different.  In Fig. \ref{fig_defect}(a), we show the final structures for $p=1, 0.99, 0.98$ and $0.97$, and the corresponding deviations $(\mathcal{O})$ from the $p=1$ mean folded structure in Fig. \ref{fig_defect}(b).

\section{Discussion and Conclusion \label{conclusion}}

In conclusion, we study how the folding pathways and mechanical stability of active elastic networks depend on the spatial distribution of active dipoles ? Our findings demonstrate that the spatial patterning of active forces is not only crucial for the mechanical stability against the contractile stresses generated by the same active forces, but also for its dynamical behavior such as folding and collapse as the activity increases. Considering two opposite cases, one in which the active dipole links are distributed like a patch at the center of the network, and along the periphery in the other case, we perform a comparative study while also comparing with the previously known results for the uniform distribution of active dipoles  from Ref. \cite{majumdar2025}.

We found that a spatially correlated cluster-like distribution enhances the overall mechanical stability of the network which undergoes only a partial decrease in size even at large active forces, which otherwise would have undergone a permanent deformation like in a collapse transition under the action of these active forces.  Further, for the two types of distribution considered, we found two different regimes of forces where one is mechanically more stable than the other; for smaller forces the active periphery is more stable while for larger forces the active core is more stable. The exponent for the scaling of network size with active patch size changes with its distribution. Changing the density of active links at a constant large force seems to  reduce the network size only linearly without any non-linear effect such as a sudden collapse.  Further, we also study the reversibility under gradual removal of forcing in these systems since taking the system back to its initial configuration after a planned task is accomplished is crucial for reusability. On the other hand, irreversibility can be of  importance when only a unidirectional process is required, e.g., during morphogenesis.  Interestingly, we found that the irreversibility depends on the distribution of active links, and the systems can be made to deform reversibly only when forces are under a threshold value, beyond which the system deforms irreversibly. For the uniform distribution this irreversibility point  lies very near to the collapse transition point making it the best choice for reusability across a wide range of forces. For active core and active periphery, however, the system responds reversibly only for a small range of forces. The irreversibility in all three cases has been linked to plastic deformation like formation of creases. While the final target state shows very weak dependence on the temporal stochastic switching of the active links, there is a very strong dependence on even an iota of defect in the network.

Our study suggests that just by spatially rearranging the active sources, e.g., the myosin motors in actin network, the overall network rigidity can be increased or decreased or can even be taken into a collapsed state without requiring any changes in the intrinsic force amplitude of the active units. This dynamic modulation  could assist cells in situations when ATP  levels (proportional to motor actvity) are low. Biology can, therefore, leverage the spatial distribution of active forces as one of the mechanisms to alter the mechanical response. Additionally, the folding of elastic sheets driven by localized active stresses could provide for a minimal physical framework to understand morphogenetic processes such as epithelial invagination and folding \cite{teranishi2024}. In our triangulated elastic network, clustered dipole forces generate inward buckling and folding, similar to the way in which apical actomyosin contractility in a group of cells drives invagination in a monolayer tissue \cite{martin2010}. Despite differences in biological complexity, both systems highlight a common physical principle: localized active stresses in a cohesive sheet can robustly trigger large-scale shape transformations.

Important to mention, our findings are not only restricted to biological systems but can be parallelly applied to other macroscopic non-biology systems as well wherever an elastic media and active force comes together, e.g, it can contribute towards understanding how active actuators can be used in mechanical elastic networks to accomplish desirable folding pathways in metamaterials. While it is clear from our study that a targeted folding can be achieved using a distribution of active links, understanding it at the level of single active links, and how a combination of few gives rise to a specific folding locally, remains a challenge. Since we work in 2D here, fold inevitably includes overlapping of network segments, however, in 3D this artefact can be avoided leading to actual fold of two-dimensional sheets. Future efforts will focus in how to improve the reversibility and therefore reusability of such networks by including additional constraints such as bond bending and excluded volume interactions which is expected to increase the overall rigidity.

\section{Acknowledgements}
DM thanks Rony Granek and Somendra M Bhattacharjee for helpful discussions. DM was supported by the BCSC Fellowship from the Jacob Blaustein Center for Scientific Cooperation.

\section{Data Availability}
The data supporting the findings is available from the corresponding author upon reasonable request.

\bibliographystyle{unsrt}
\bibliography{networks.bib}

\begin{thebibliography}{10}
\bibitem{comm1}
Maxwell's rule for mechanical stability states that in a $d$-dimensional
  network at the isostatic point the mean coordination number $\langle z
  \rangle$ should be at least $\langle z \rangle=2d$ which is 4 for
  two-dimensional systems.
\end{thebibliography}

\appendix

\begin{figure}[h]
\centering
\includegraphics[width=.95\linewidth]{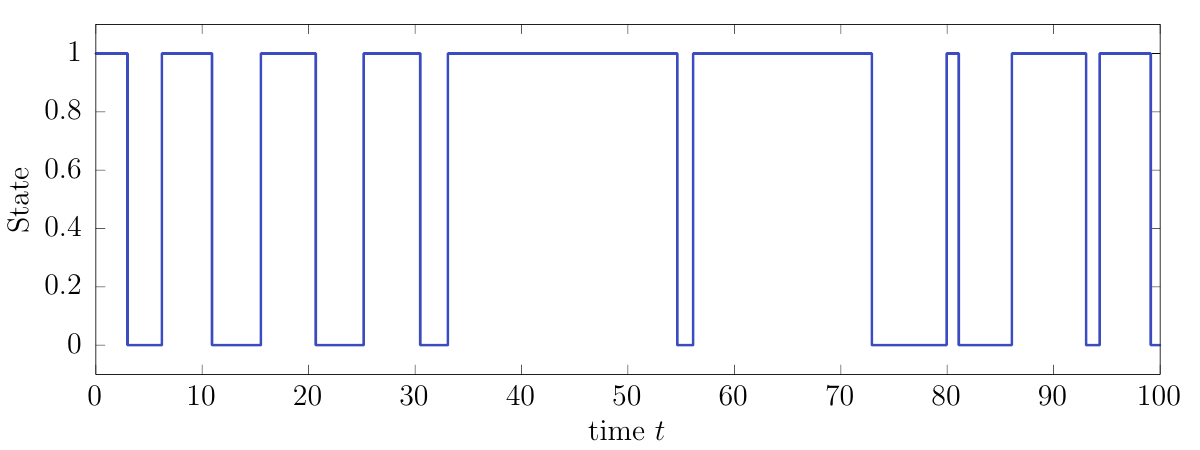}\\
\includegraphics[width=.47\linewidth]{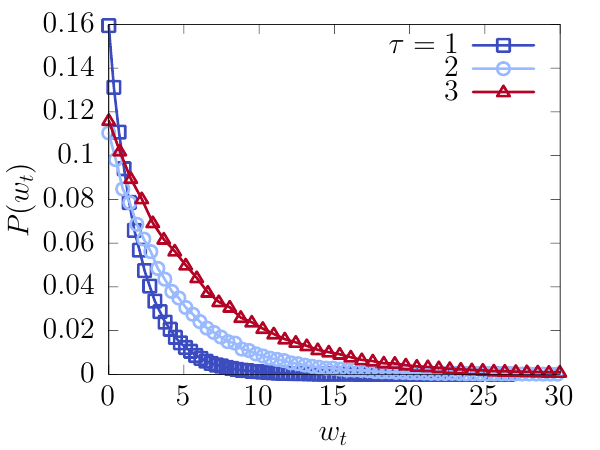}
\includegraphics[width=.47\linewidth]{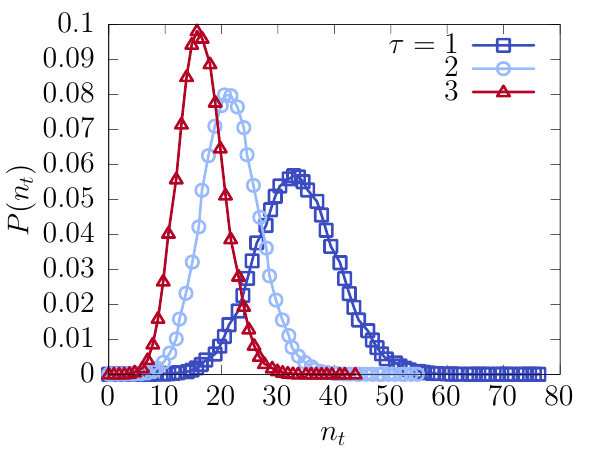}
\caption{\textbf{Transition statistics of a force dipole.} (a) Time series plot of a force dipole switching between ON $(1)$ and OFF $(0)$ states, shown for $\tau=3$ upto time $t=100$. (b) Here, we show the exponential distribution of the waiting $(P(w_t))$ times before the active dipole changes state. (b) Probability distribution of number of transitions $(P(n_t))$ made within a time interval $t=10$ follows a Poissonian distribution.}
\label{fig_active_dipole_stat}
\end{figure}

\section{Switching statistics of force dipoles \label{appendix_active_force}}

The active links are modelled using telegraphic noise with a finite decorrelation time $\tau$. Within this time scale, the switching of active links between the ON and OFF states remains temporally correlated. The stochastic jumping between the states $0$ and $f$ follows the following probability equations

\begin{eqnarray}
 R &\leq&  (1-{P})+{P}\times \exp(-\delta t/\tau),~~~~ 0\rightarrow 0 \\
 R &\leq& {P}+(1-{P})\times \exp(-\delta t/\tau),~~~~ f\rightarrow f,
\end{eqnarray} 

where $P$ is the probability to be in the ON state during which the active dipoles contract the spring bonds with force $f$,  $R\in[0,1]$ is a uniform random number and $\delta t=10^{-2}$ is the integration time step. In Fig. \ref{fig_active_dipole_stat}(a), we show the distribution $P(w_t)$ of the intermittent waiting time lengths $(w_t)$ for which the system stays in one of the state before it jumps into the other state which follows an exponential distribution, i.e., $P(w_t)\sim exp(-w_t/\tau')$, where $\tau'\propto \tau$. In Fig. \ref{fig_active_dipole_stat}(b), we show that the number of transitions $(n_t)$ made within a time interval $t=100$ and averaged over $10^8$ samples, follows a Poissonian distribution of the form $P(n_t)= \frac{\lambda^{n_t}e^{-\lambda}}{(n_t)!}$, where $\lambda\propto 1/\tau$.

\section{Integration scheme \label{appendix_int_scheme}}

To numerically integrate the ordinary differential equation (ODE) of motion we use the two-point finite difference method, also known as the second order Heun algorithm or the midpoint scheme for ODEs. This consists of the conventional Euler-Maruyama scheme (Eq. \ref{euler}(a)) along with an additional corrector step (Eq. \ref{euler}(b)) as mentioned below 
\begin{subequations}
\begin{eqnarray}
\bar{\vec r}_{k+1} &=& \vec r_k + \frac{1}{\gamma}\vec F_e (\vec r_k) \delta t  +  \vec F_a(\vec r_k) \delta t   \\
\vec r_{k+1} &=& \vec r_k + \frac{1}{2\gamma} \left( \vec F_e (\vec r_k) + \vec F_e (\bar{\vec r}_{k+1}) \right) \delta t  \nonumber \\ 
&+&   \vec F_a(\vec r_k) \delta t\end{eqnarray}
\label{euler}
\end{subequations}

where $\vec F_e (\vec r_k)$ are the total elastic forces and  $\vec F_a(\vec r_k)$ are the active forces. In this integration scheme, a predicted position $\bar{\vec r}_{k+1}$ (Eq. \ref{euler}a) is followed by a corrected position $\vec r_{k+1}$ (Eq. \ref{euler}b). Notice that the same active force components  $\vec F_a$ are to be used both in the predictor and corrector step and the averaging is performed only upon the spring (conservative) forces.

\section{On the scaling exponents \label{appendix_scaling_exponents}}

To understand the origin of the exponent in the scaling of $R_g$, in the high force regime, for the active core network, we notice that the number of nodes in $d$ dimension scales as $N\sim L^d$, where $L$ is the linear dimension of the network, therefore, $R_g\sim L$. Then, we can write $R_g\sim N^{1/d}$. For the active core, after the core collapses to a size $\sim b$, the maximum contribution to $R_g$ comes from the passive nodes only. The number of passive nodes excluding the active core is roughly $N'\sim(1-\phi)N$. Then, in the strong force regime after the active core has collapsed, $R_g'\sim N'^{1/d}\sim (1-\phi)^{1/d}$. Since, $1/d=0.5$, there is still a factor of $2$, which we believe comes from the specific distribution of the active force dipoles.

\end{document}